\documentclass{nature-preprint}
\usepackage{tcolorbox,color}
\usepackage{colortbl}
\usepackage{ulem}
\usepackage{amsmath}
\usepackage{amssymb}
\usepackage{graphicx}
\usepackage{upgreek}

\usepackage{caption,wrapfig}
\usepackage{hyperref}
\usepackage{lipsum}
\title{Giant enhancement of silicon plasmonic SWIR photodetection using nanoscale self-organised metallic films}

\author{Christian~Frydendahl$^{1,*}$, Meir~Grajower$^{1}$, Jonathan Bar-David$^{1}$, Noa~Mazurski$^{1}$, Joseph~Shappir$^{1}$, and Uriel~Levy$^{1,*}$}

\begin{document}

\thispagestyle{empty}
\includegraphics[width=\columnwidth]{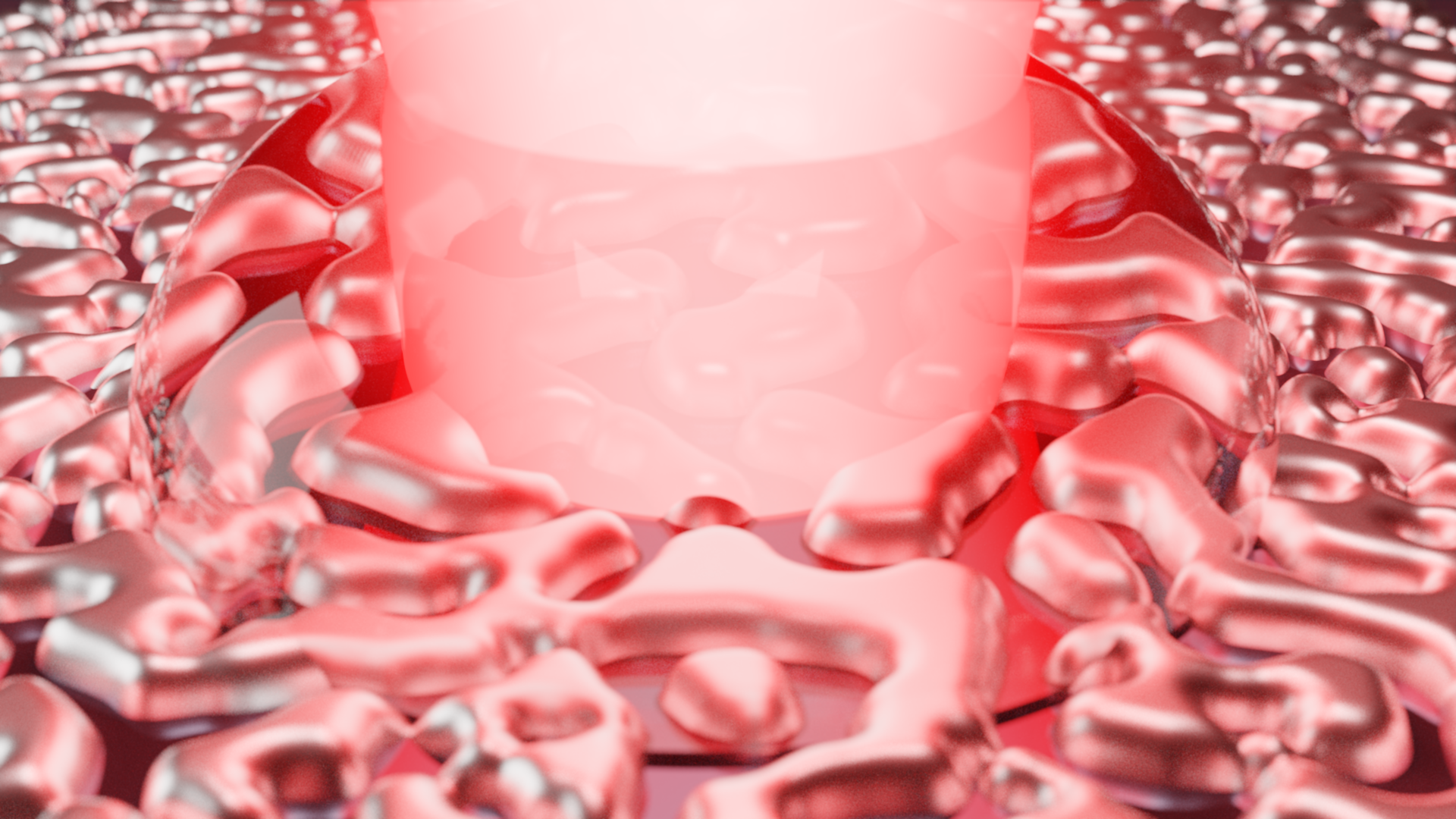}\\

\newpage
\setcounter{page}{1}
\maketitle % this command generates the title and author list

{\linespread{1}
% the affiliations
\begin{affiliations}
\item  Department of Applied Physics, The Faculty of Science, The Center for Nanoscience and Nanotechnology, The Hebrew University of Jerusalem, Jerusalem, 91904, Israel
\item[]
\item[*] christia.frydendahl1@mail.huji.ac.il, ulevy@mail.huji.ac.il
\end{affiliations}
}

%%%%%%%%%%%%%%%%%%%%%%%%%%%%%%%%%%%%%%%%%%%%%%%%%%%%%%%%%%%
%here goes the abstract
% max 150 word

\begin{abstract}
Many consumer technologies and scientific methods rely on photodetection of infrared light. We report a Schottky photodetector operating below silicon's band gap energy, through hot carrier injection from a nanoscale metallic absorber. Our design relies on simple CMOS-compatible 'bottom up' fabrication of fractally nanostructured aluminium films. Due to the fractal nature of the nanostructuring, the aluminium films support plasmonically enhanced absorption over a wide wavelength range. We demonstrate two orders of magnitude improvements of responsivity, noise-equivalent-power, and detectivity as compared to bulk metal, over a broad spectral and angular range. We attribute this to momentum relaxation processes from the nanoscale fractal geometry. Specifically, we demonstrate a direct link between quantum efficiency enhancement and structural parameters such as perimeter to surface ratio. Finally, our devices also function as bulk refractive index sensors. Our approach is a promising candidate for future cost effective and robust short wave infrared photodetection and sensing applications.
\end{abstract}

% word count = 150 words.

%%%%%%%%%%%%%%%%%%%%%%%%%%%%%%%%%%%%%%%%%%%%%%%%%%%%%%%%%%%
\section*{Introduction}
Photodetection of near infrared (NIR) and short wave infrared (SWIR) light is of interest for a large variety of practical applications, such as optical communication\cite{Keiser:2003}, spatial proximity sensing\cite{Blais:2004}, night vision\cite{Chrzanowski:2013}, as well as molecular spectroscopy - which itself has applications ranging from topics like medicine\cite{Parker:2012} and food science\cite{Osborne:2000}, to astrobiology\cite{Schwieterman:2018}. As such, there is a large desire for cheap, efficient, and scalable technologies for fabrication of NIR and SWIR photodetectors. However, silicon, the staple of complementary metal-oxide-semiconductor (CMOS) technologies, becomes transparent in the NIR, as the band gap of silicon is roughly 1.1\,eV ($\sim$1.1\,$\upmu$m wavelength). Thus silicon is not the natural choice for photodetection of wavelengths above $\sim$1.1\,$\upmu$m.

A Schottky junction between a metal and semiconductor can allow for optical absorption of sub-band gap light in the metal, and correspondingly photodetection of photons below the semiconductor band gap, through a process known as internal photo-emission\cite{Mishra:2007,Scales:2010,Berini:2017}. After a photon is absorbed in the metal, and a hot carrier is generated, it is possible for this carrier to emit across the Schottky junction barrier, $\Phi_\text{B}$\cite{Scales:2010,Berini:2017,Levy:2017,Grajower:2018a}. This injection of hot carriers into the semiconductor can then be collected as a photocurrent, which is proportional to the incident intensity of light on the junction. The proportionality factor, $R$, is then known as the photodetector's responsivity. Schottky junctions thus allow semiconductors, like silicon, to indirectly absorb light with energies that exceed $\Phi_\text{B}$, which can be considerably below the band gap energy of the bulk semiconductor. The magnitude of $\Phi_\text{B}$ depends on the choice of metal and semicondutor, as well as the doping of the semiconductor\cite{Berini:2017} and its surface quality. Typical values of $\Phi_\text{B}$ for metal/silicon junctions, will range from 0.2 to 0.8\,eV\cite{Mishra:2007,Scales:2010,Berini:2017}. 

However, a common issue with Schottky photodetectors is their low responsivities (typically $R \sim 1$\,$\upmu$A/W). Various schemes have been developed to increase the optical absorption efficiency of the metal, and secondly the efficiency of the transfer of the hot carriers to the semiconductor\cite{Scales:2010,Berini:2017,Levy:2017}. Such schemes consist of optical waveguide structures\cite{Akbari:2009,Goykhman:2011,Goykhman:2012}, plasmonic absorbers - such as antenna\cite{Knight:2011,Alavirad:2013,Li:2014,Li:2015,Desiatov:2015} and grating arrays\cite{Brueck:1985,Sobhani:2013,Giugni:2013,Alavirad:2016,Lin:2014}, or random nanostructures\cite{Lee:2011,Wen:2017}. Engineering of the metal thickness in the junction\cite{Scales:2010}, and optimisation of the optical coupling in the device by cavity modes\cite{Casalino:2012,Kosonocky:1985}, has also proven successful. Recently, the addition of an intermediate layer of graphene between the metal and semiconductor, has also been shown to drastically increase the transfer efficiency of generated hot carriers\cite{Goykhman:2016,Levy:2017}. The role of surface roughness in the interface has also recently been studied\cite{Grajower:2018a}, and combined Schottky-photoconductor devices with gain have also been studied\cite{Tanzid:2018}.

Metal percolation films (also known as semi-continuous films or metal-dielectric composite films) have previously been noted for their high optical absorption\cite{Dobierzewska:1985,Yagil:1987,Kunz:1988,Yagil:1992,Berthier:1997}. This is typically attributed to the high degree of plasmonic field enhancement found in such films\cite{Ducourtieux:2000,Genov:2003,Shalaev:2007}. Optical fields can couple to collective oscillation modes of the conduction electrons at the surface of metallic nanostructures, known as plasmons\cite{Schuller:2010}. Exciting plasmons results in strongly localised electrical fields that can be several orders of magnitude higher than that of the initial incident optical field, resulting in stronger light-matter coupling, and these plasmons can also decay directly into hot carriers\cite{Schuller:2010}. Percolation films consist of a fractal network of random metallic clusters, containing gaps and metallic particles in a wide range of sizes, ranging from sub-nanometer to hundreds of nanometres\cite{Shalaev:2007,Frydendahl:2017}. This is what gives these structures their very broad spectral range of optical enhancement\cite{Shalaev:2007,Genov:2003}. 

Percolation structures can be achieved from simple metal evaporation processes, as their formation relies on the intrinsic growth patterns of metals on dielectric substrates during deposition\cite{Malureanu:2015}. Because of the surface energy between metals and dielectrics, the formation of a thin metal film will occur based on a 3D island growth pattern, also known as Volmer-Weber growth\cite{Malureanu:2015}. As metal is deposited, the metal adatoms will have a preference to 'stick' to other metal adatoms, rather than the substrate. This results in initial seeding into separated metallic islands in the early stages of deposition. With increased deposition of metal, the isolated islands will slowly grow laterally across the substrate, until they eventually start to connect together and form a dominant 'super cluster'. This point is known as the percolation threshold, and marks the point where the thin metal film's properties become dominated by the one unified cluster of metal\cite{Shalaev:2007}. For example, the percolation film becomes electrically conductive at the percolation threshold, as it is now possible to find an electrically connected path from one side of the film to the other (however windy or tortuous). Additional metal deposition after this point only serves to close any remaining gaps in the film, and will eventually result in full metal coverage of the substrate, forming a smooth metal film\cite{Malureanu:2015}. Metal percolation films have been a popular topic of research in past decades, due to their high intrinsic optical field enhancement, and ease of fabrication. They have previously been used for such applications as surface enhanced Raman scattering (SERS) substrates\cite{Gadenne:1997, Drachev:2005,Perumal:2014, Novikov:2016, Wallace:2019}, extinction/absorption enhancement\cite{Galanty:2015}, enhancement of non-linear optical processes in gold, such as two-photon luminescence\cite{Ducourtieux:2000,Novikov:2016} and white-light generation\cite{Ducourtieux:2000,Novikov:2017,Chen:2018}. Recently, they have also been included in novel plasmonic laser printing schemes\cite{Novikov:2017,Frydendahl:2017,Zheng:2017,Roberts:2018}.

Here we present a simple method for high-efficiency sub-band gap photodetection in silicon, using a cheap, CMOS-compatible, and highly scalable fabrication technique, eliminating the need for costly and complicated nanoscale lithographic processes. By forming a Schottky contact between a p-doped silicon substrate and fractally nano-patterned aluminium metal percolation films, we are able to guarantee high optical responsivities across a wavelength range of 1304\,nm to 1550\,nm, with peak responsivity of $R\sim4.5$\,mA/W and peak internal quantum efficiency of $\eta_i\sim1.0\%$ both at a wavelength of 1304\,nm. In addition, we also demonstrate how the absorption in our devices is sensitive to the bulk refractive index surrounding the metal films, because of changes to the optical coupling efficiency. This allows the devices to function as refractive index sensors for liquids placed on top of the metal films, with a direct electrical read-out. This establishes a new area of Schottky photodetector applications in the chemical- and biosensing domains\cite{Alavirad:2013,Tsukagoshi:2018}, beyond just cost-effective short wave photodetectors.

%%%%%%%%%%%%%%%%%%%%%%%%%%%%%%%%%%%%%%%%%%%%%%%%%%%%%%%%%%%
\section*{Results}
\subsection{Sample design:}
\begin{figure}[h]
	\centering
	\includegraphics[width=15cm]{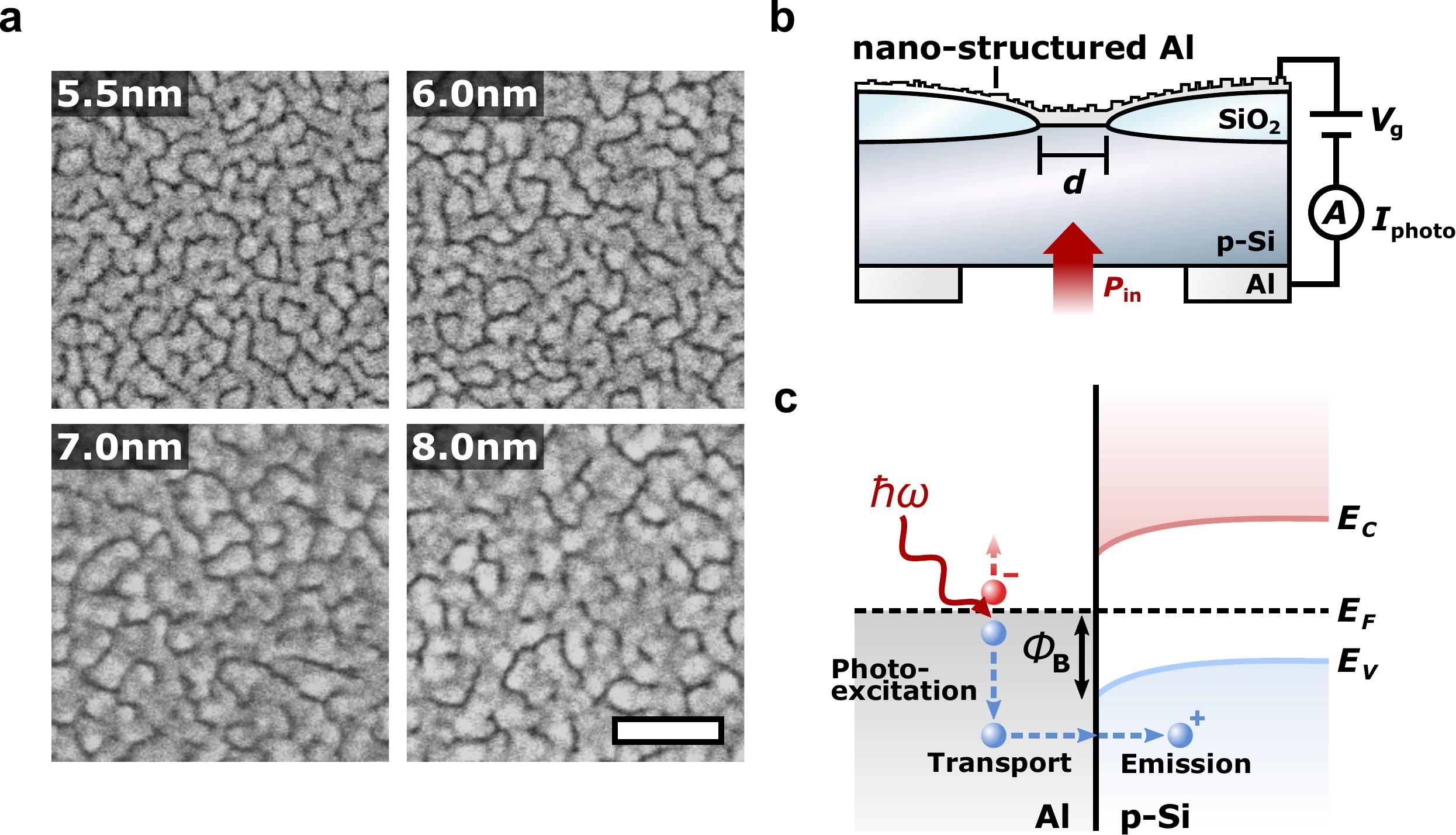}
    \caption{Overview of investigated samples. \textbf{a)} SEM micrographs of the nanostructures achieved at the diode areas for different metal deposition thicknesses. All images at same scale, scale bar is 125\,nm. \textbf{b)} Schematic side view of the diode geometry, showing the sloped side walls of the insulating oxide. The illumination scheme is shown with $P_\text{in}$. The applied bias, $V_\text{g}$, across the junction and the measured photocurrent, $I_\text{photo}$, are also defined. The active diode area is $d^2 = 40 \times 40$\,$\upmu$m$^2$. \textbf{c)} The internal photo-emission process of hot holes into silicon. After optical absorption in the aluminium of a photon with energy $\hbar\omega>\Phi_\text{B}$, there is a chance to emit the hot hole into the silicon valence band, resulting in a photocurrent.}
    \label{fig:SEM}
\end{figure}

We have fabricated 5 devices, differing only in the amount of aluminium deposited for the top electrode forming the Schottky junction to the p-doped silicon substrate. The samples have nominal deposited thicknesses of 5.5, 6, 7, and 8\,nm. All of these thicknesses are above the percolation threshold\cite{Novotny:2011}, and thus electrically conductive. A device with a 75\,nm bulk aluminium electrode was also fabricated to act as a control sample. All studied diodes have square active areas of $40 \times 40$\,$\upmu$m$^2$. Scanning electron microscope (SEM) images of the nanostructures generated by varying the aluminium deposition, can be seen on Fig.~\ref{fig:SEM}.a.

\linespread{1}

In the sample fabrication, we have utilised the technique of local oxidation of silicon (LOCOS) to define the square diode windows, see Fig.~\ref{fig:SEM}.b. This ensures a smooth slope of the SiO$_2$ sidewalls\cite{Goykhman:2011}, allowing the thin aluminium films to only touch the exposed silicon in a small area, and without breaking the electrical connection of the film from a steep change in height across the sample. For the full details of the sample fabrication, see methods and Supplementary Fig.~1. As we illuminate the samples from below, through the silicon, the top surface of the percolation film electrode is left freely accessible to introduce liquid droplets for sensing applications.

A simple schematic outline of the internal photo-emission process can be seen in Fig.~\ref{fig:SEM}.c. The process consists of 3 steps. First, when a photon of an energy $\hbar\omega$ is absorbed in the aluminium, a hot electron-hole pair is generated. The hole will then propagate inside the metal with a random momentum. If the energy of this hole exceeds the Schottky barrier of $\Phi_\text{B}$, and the momentum of the hole matches to the silicon, it is possible to emit the hole into the p-type silicon's valence band. Here it will contribute to a measurable photocurrent. The efficiency of this emission process defines the overall internal quantum efficiency, $\eta_{i}$ (and in turn, the responsivity $R$), of a Schottky photodiode operating in the sub-band gap regime. It has generally been reported, that making the metal electrode in the Schottky junction thinner than the thermalisation length of the carriers in the metal, results in a large increase in responsivity\cite{Scales:2010,Berini:2017,Li:2014}. This is because of two reasons: 1 - there is higher probability for the carriers to arrive at the metal-semiconductor interface before undergoing thermalisation, and 2 - even if the hot carrier is not emitted when encountering the Schottky barrier for the first time (due to a momentum mismatch when crossing between the two materials\cite{Scales:2010}), it can be reflected back and forth between the thin metal film's interfaces. This gives the carrier multiple attempts to cross the Schottky barrier, before it is attenuated inside the metal and loses its energy\cite{Scales:2010,Levy:2017}. As the films in our structures are all in the percolation regime, we can safely expect that the clusters of metal are generally only a few nanometres thick, i.e. smaller than the thermalisation length.

The second important aspect in our percolation structures is the effect of plasmonic field enhancement. Due to the self-similar/fractal nature of the nanostructuring in percolation films, it has been shown previously that they support plasmonic 'hotspots' in a wide range of optical frequencies\cite{Genov:2003,Shalaev:2007,Frydendahl:2017}. The effect of such plasmonic hotspots is generally to confine any incident optical fields close to the surface of the metal, increasing the local field intensity by several orders of magnitude\cite{Genov:2003,Novikov:2017}. By confining and concentrating optical fields down to the nanoscale, it is possible to greatly increase light-matter interaction - such as optical absorption into the aluminium electrodes. Metal percolation films have previously been reported to show absorption of around 20 to 30\% in visible and NIR\cite{Novikov:2016,Kunz:1988,Pribil:2004,Berthier:1997}. The expectation then, is a large increase in the device's external quantum efficiency, $\eta_e$, when compared to a bulk film. The general relationship between the internal and external efficiencies is $\eta_e = A \eta_i$, with $A$ being the absorption of the device\cite{Scales:2010}. The responsivity of a device is then likewise given as\cite{Scales:2010}:
\begin{align}
	R(\omega) = \frac{A \eta_i e}{\hbar\omega},
	\label{eq:R}
\end{align}
with $e$ as the elementary charge. From this we see that it is essential to increase the absorption of a device, as well as its internal quantum efficiency, to insure the highest levels of responsivity.

The noise equivalent power (NEP), is the lowest amount of incident power that a detector can detect at a signal to noise ratio (SNR) of 1, at an output bandwidth of $\Delta f = 1$\,Hz:
\begin{align}
    \text{NEP}(\lambda) = \frac{\sqrt{2 e I_0}}{R(\lambda)},
\end{align}
With $I_0$ as the dark/leakage current. Likewise, the specific detectivity of a detector, $D^*$, is then defined by the detector's area, $A_\text{det}$, and the output bandwidth (integration time), as:
\begin{align}
    D^*(\lambda) = \frac{\sqrt{\Delta f A_\text{det}}}{\text{NEP}(\lambda)}.
\end{align}
The NEP of a detector thus helps define its sensitivity more accurately than the responsitivity alone, and the specific detectivity is an even more precise figure of merit for a detector, as it accounts for the total active area of the device, which is likely to affect the dark current and noise.

\subsection{Optical responsivity and quantum efficiency:}

\begin{figure}[h]
	\centering
	\includegraphics[width=12cm]{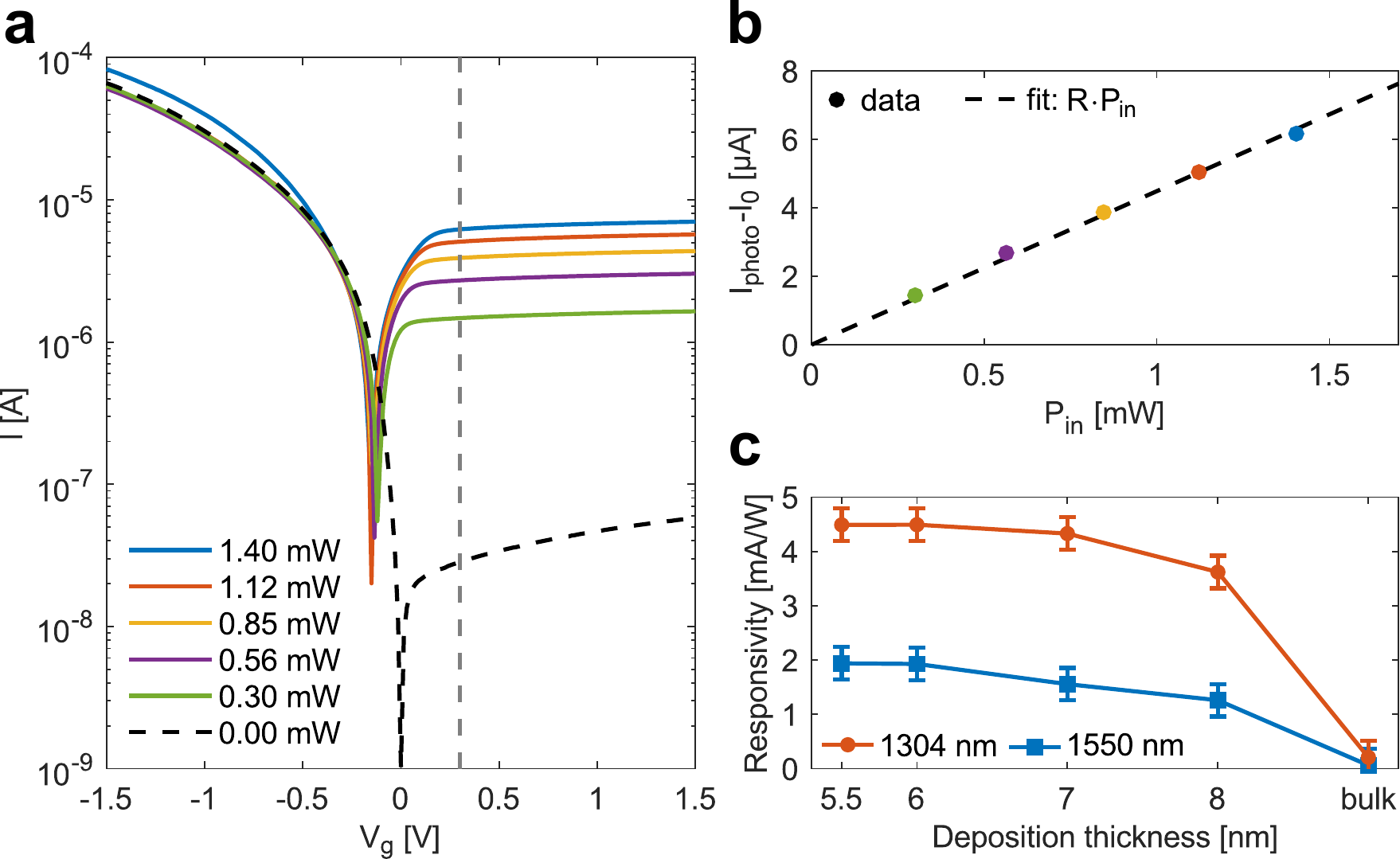}
    \caption{Responsivity characterisation. \textbf{a)} Example of $IV$-characteristics of a 5.5\,nm thick film diode, with and without various powers of 1304\,nm illumination. The dotted vertical line marks the 0.3\,V reverse bias used for the responsivity calculations in \textbf{b-c}. \textbf{b)} Example of the linear regression for responsivity determination of the 5.5\,nm device in \textbf{a}, at 0.3\,V reverse bias. \textbf{c)} Responsivity vs. Al deposition thickness, for two different wavelengths of excitation. All responsivities are calculated for $V_\text{g}=0.3$\,V. A bulk film (75\,nm) has also been included for comparison. Error bars are multiplied by 6 for better visualisation.}
    \label{fig:Res}
\end{figure}

We investigated the spectral dependence of our devices' responsivities for wavelengths of 1304 and 1550\,nm (see methods), the results can be seen on Fig.~\ref{fig:Res}. The devices are all illuminated from below through the silicon, as indicated in Fig.~\ref{fig:SEM}.b. An example $IV$-measurement from a diode with a 5.5\,nm Al deposition can be seen on Fig.~\ref{fig:Res}.a. By recording the generated photocurrent at different incident optical powers, we can determine the responsivity from a simple linear regression, see Fig.~\ref{fig:Res}.b.

\linespread{1}

From Fig.~\ref{fig:Res}.c, we find generally high responsivities in the range of $R \sim 1-4.5$\,mA/W for all 4 investigated percolation film thicknesses. This is 2 orders of magnitude higher than the responsivity of our bulk control sample (a 75\,nm thick film), which is of the order $R_\text{bulk} \sim 0.01$\,mA/W.

There is no appreciable increase in dark current or noise from our percolation devices when compared to the bulk device. For the 5.5\,nm devices, we get a NEP of $\sim$$2.70 \cdot 10^{-11}$\,W/$\sqrt{\text{Hz}}$, and a $D^*$ of $\sim$$1.48 \cdot 10^{8}$\,Jones for 1304\,nm light, when operating at a 0.3\,V reverse bias. For the full NEP and $D^*$ results for all devices and wavelengths, see Supplementary Fig.~2.

We find no strong dependence of the responsivity on the polarisation of the incident light. This is also to be expected, as the nanostructuring of the percolation films is random/isotropic. Although the exact spatial distribution of plasmonic hotspots will be different for different polarisations of light\cite{Novikov:2017}, the average amount and intensity of the hotspots will be similar for two mutually perpendicular polarisations. Due to the quite large refractive index of silicon ($n_\text{Si}\sim$3.5), any illumination from below through the silicon is going to experience a very high degree of refraction. This naturally makes the device less sensitive to the incidence angle of illumination, and we indeed do not see any major shift in responsivity between $0^\circ$ and $25^\circ$ of free-space angles of incidence, see Supplementary Fig.~3.

\subsection{Bulk refractive index sensing:}
\begin{figure}[h]
	\centering
	\includegraphics[width=13cm]{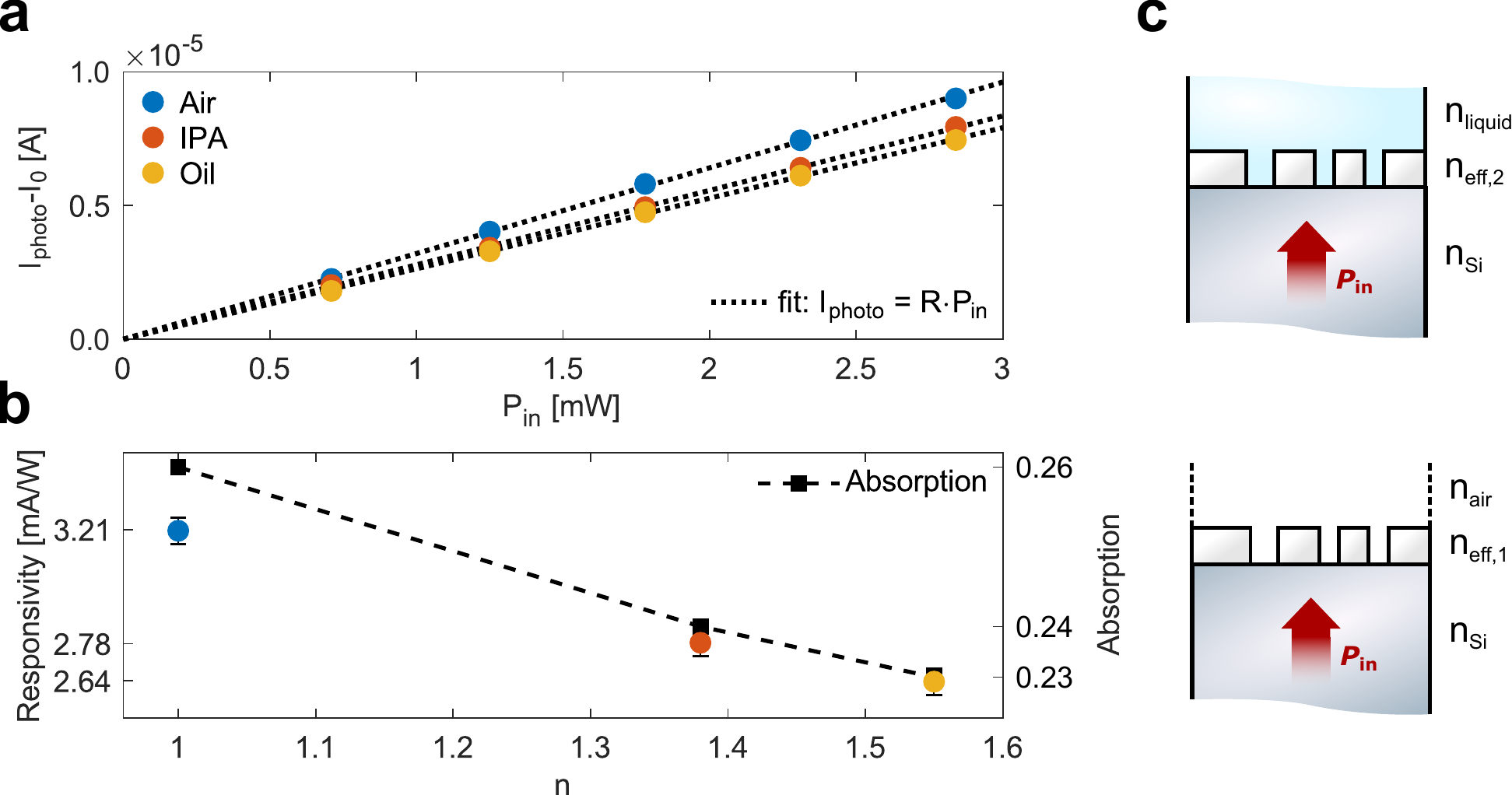}
    \caption{Liquid sensing experiments. \textbf{a)} Photocurrent of a 6\,nm film device in air and with different liquid droplets added on top, as a function of the incident optical power. Dotted lines show fits for extracting $R$. \textbf{b)} Responsivities of \textbf{a} as a function of liquid refractive index, and calculated thin film absorption (see main text). Both \textbf{a} and \textbf{b} were recorded with 1304\,nm excitation. \textbf{c)} Fresnel equation thin film model. The effective index for the metal film is calculated from the metal filling fraction and the surrounding refractive index.}
    \label{fig:sens}
\end{figure}

Plasmonic systems have previously been exploited for sensing purposes to great success in the past\cite{Stewart:2008}. However, such sensing devices are typically reliant on a very complex optical read-out of the changes in the plasmonic resonance. Here, we can directly monitor the absorption changes in the aluminium percolation film from the measured photocurrent. The results of our bulk sensing tests can be seen on Fig.~\ref{fig:sens}.a and b.

\linespread{1}

We tested a 6\,nm device using two different liquids, a 95\% isopropyl alcohol (IPA) solution ($n_\text{IPA} \sim 1.38$) and Nikon microscope immersion oil Type A ($n_\text{oil} \sim 1.52$). From Fig.~\ref{fig:sens}.a and b, we see that the responsivity of the device goes down when a higher index liquid is placed on top.

We attribute the loss in responsivity to the change in effective index of the metal film. As the metal is porous the liquid can fill out the voids in the film, and this changes the coupling efficiency to the metallic layer\cite{Krayer:2017}. Likewise, any Fabry-P\'erot cavity effect will be lessened as the higher index liquids act as an anti-reflection coating, again lowering the total coupling efficiency to the metallic absorber. We can calculate the absorption in the metal film layer as a thin coating on the silicon, using the complex Fresnel equations\cite{Krayer:2017} (model shown in Fig~\ref{fig:sens}.c). Because the metal film is porous, we use the Maxwell-Garnett equation to calculate the effective refractive index between the mixing of the metal and the bulk dielectric. Each added liquid thus changes the effective index of the metallic layer, and results in a change in total absorption, seen in Fig.~\ref{fig:sens}.b. See the supplementary materials for the full calculation details. It should be noted however, that the true absorption in the metal films is higher, as neither the Fresnel nor the Maxwell-Garnett equations consider the possibility of plasmonic resonances. The result presented here is thus only an estimate, but the relative shifts in absorption predicted by the model, and the overall trend should still apply to our case.

An alternative explanation could be additional loss mechanisms opening up from the increased spill-out of the electron density of the metal, with increasing surrounding bulk index\cite{Liu:2015}. Although our devices appear to be in the Drude-loss dominated regime, due to the relatively low refractive indices tested, and from the fact that we are working with infrared resonance frequencies\cite{Liu:2015}, the extreme nanoscale dimensions of the films might also make them more sensitive to quantum effects\cite{Christensen:2017}.

Because our percolation film electrode structure hosts a wide continuum of resonances, the sensing is not based on the redshifting of resonances from adding a higher index material on top of the film, as the density of resonances is expected to stay similar within a narrow shift of wavelengths\cite{Frydendahl:2017}.

\subsection{Fractal properties of the film geometries:}

We investigate the features of the different film geometries by using image analysis of SEM images recorded of each of the film geometries. Because all of our samples are above the percolation threshold, we cannot easily study the properties of the metal itself, as there is only one connected component of metal. We choose instead to study the inverse geometry, and investigate the absences of metal in the films, i.e. the voids. The full results of this analysis can be seen in the supplementary materials (Supplementary Figs.~4-6), but the overall conclusions are as follows:

\linespread{1}
\begin{figure}[h]
	\centering
	\includegraphics[width=\columnwidth]{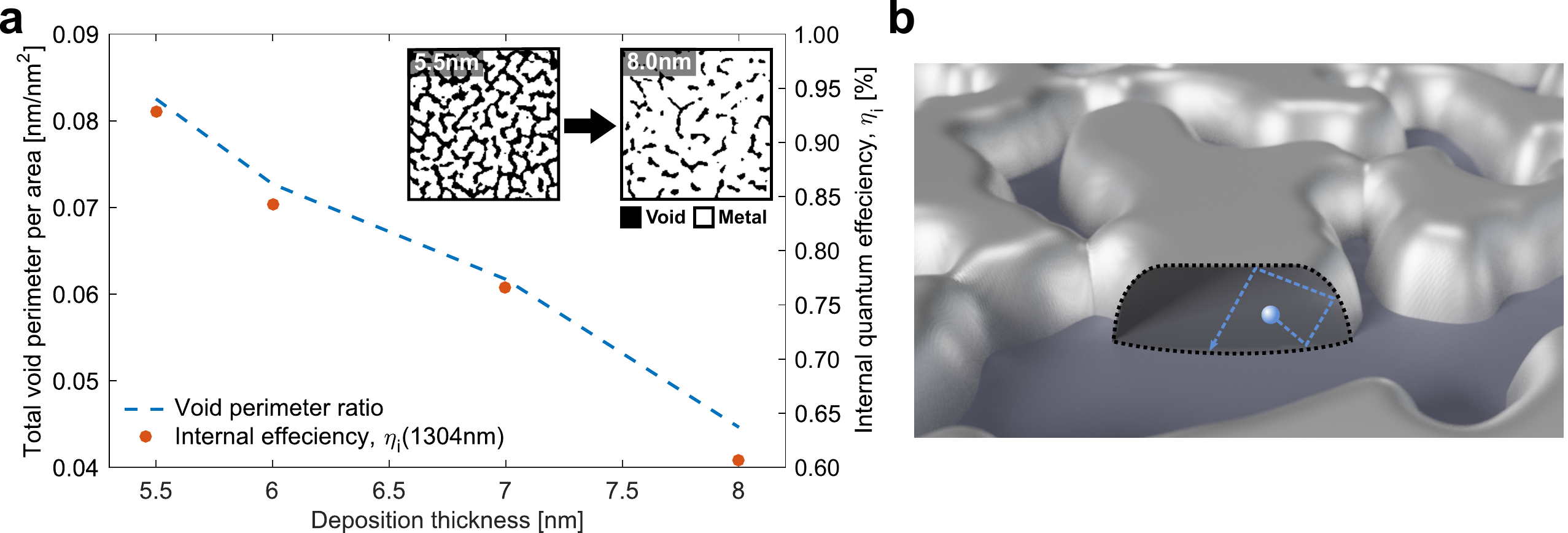}
    \caption{Momentum relaxation. \textbf{a)} Total perimeter of voids normalised to the area investigated (dotted line), and internal quantum effeciencies, $\eta_i$ for 1304\,nm (points), for the different film geometries. The inset shows a binary image of a 5.5\,nm film as compared to an 8.0\,nm film. \textbf{b)} Schematic of how the localisation of carriers in the nanoscale features in the films results in momentum relaxation, easing the transmission process of carriers from the metal to the silicon.}
    \label{fig:frac}
\end{figure}

The thinner metal depositions results in fewer, but significantly bigger and more complex void shapes. This is shown in the supplementary materials by the void clusters of the thin films having significantly larger correlation lengths, $\xi$, average areas, $S$, and higher fractal Hausdorff dimensions, $D$. As shown in Fig.~\ref{fig:frac}, the thinner structures are also significantly more dominated by the perimeter of voids per the total film area (i.e. the images contains more metallic edges). This means that corresponding metallic clusters are more separated into thin filaments, rather than a continuous film with small isolated voids as in the thicker depositions (see Fig.~\ref{fig:frac}.a inset).

We determine the internal quantum efficiency of our devices from absorption measurements performed with an integrating sphere, see Supplementary Fig.~7. From these measurements, we see that absorption goes from roughly 0.45 to 0.55 from the thinnest to the thickest films. This is likely due to enhanced coupling efficiency to the thicker films, from the Fabry-P\'erot effect discussed in the section above. Despite the increasing absorption, we see that the responsivity drops for the thicker films in Fig.~\ref{fig:Res}. To explain this, we calculate the internal quantum efficiencies of the films using the measured responsivities and absorption values for 1304\,nm, and follow eqn.~\ref{eq:R}. Fig.~\ref{fig:frac}.a shows the found internal quantum efficiencies. The peak value is of roughly 1\% for the 5.5\,nm film devices. However, what is extremely interesting, is how there seems to an almost direct relationship between the change in quantum efficiency and the total void perimeter ratio of the films (i.e. how much of a film image is made up of metallic edges). This is also mirrored for the case of 1550\,nm, shown in Supplementary Fig.~8

Films with a higher total void perimeter ratio result in more isolation of the metal into thin filaments and nanoclusters. This isolation enables a higher degree of hot carrier momentum relaxation, as the carriers are more likely to encounter a metal/vacuum interface and elastically scatter off of it to change their momentum, before the carriers can thermalise. See Fig.~\ref{fig:frac}.b for a schematic of the process.

\subsection{Schottky barrier measurements:}
\begin{figure}[h]
	\centering
	\includegraphics[width=12cm]{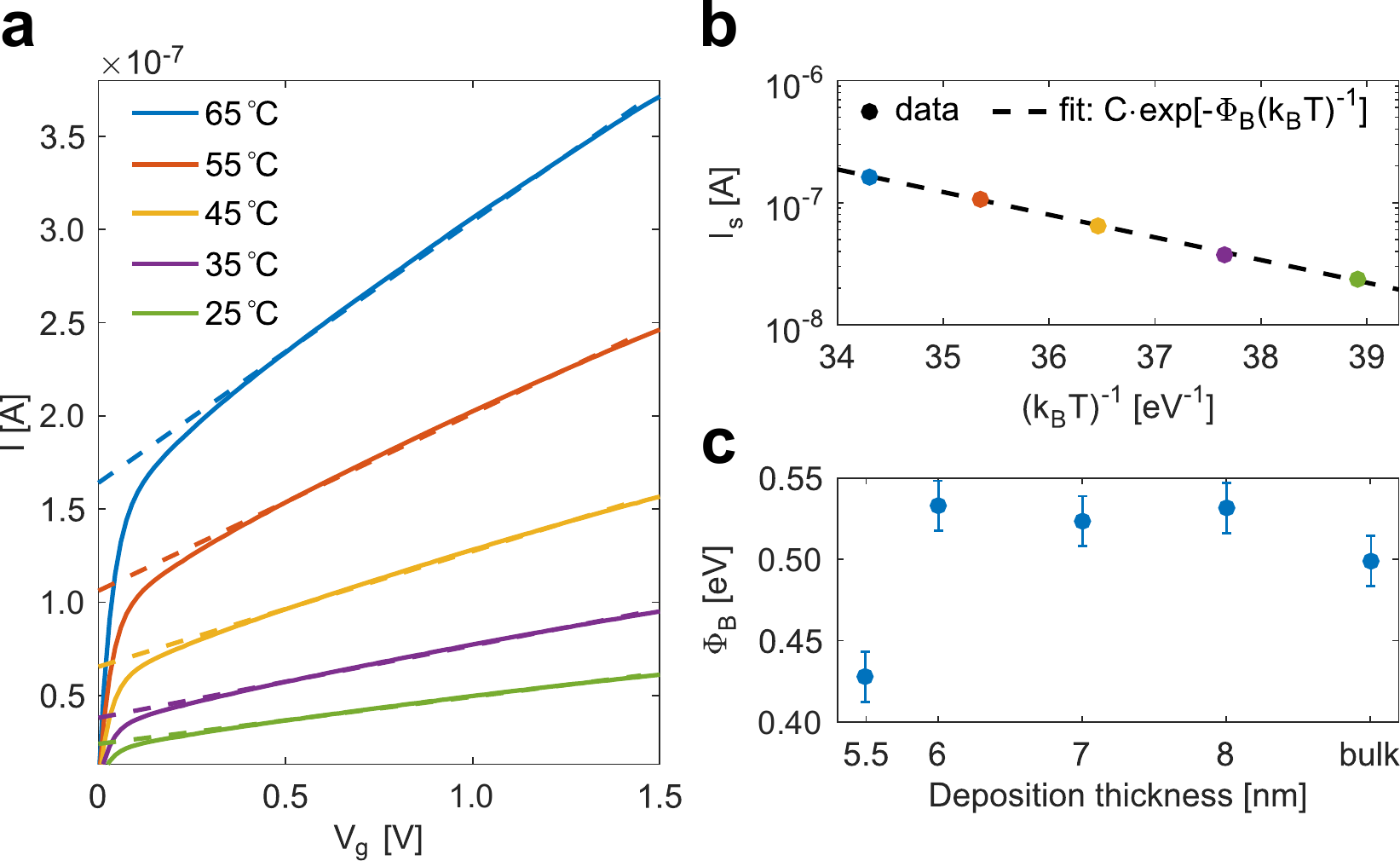}
    \caption{Schottky barrier characterisation. \textbf{a)} Temperature dependent IV-curves for a 5.5\,nm device in reverse bias. The dotted lines indicate linear regressions for extracting the saturation current, $I_\text{s}$, which is used for determining the Schottky barrier (see main text/methods). \textbf{b)} Exponential fit for determining the Schottky barrier height of the 5.5\,nm device highlighted in \textbf{a}, using the extracted values of $I_\text{s}$. \textbf{c)} Comparison of $\Phi_\text{B}$ determined for devices of different deposition thicknesses.}
    \label{fig:Sch}
\end{figure}

Finally, we have characterised the Schottky barrier height in our different devices (see methods). The results can be seen on Fig.~\ref{fig:Sch}.

\linespread{1}

Fig.~\ref{fig:Sch}.a shows an example of the temperature dependent $IV$-characteristics of a 5.5\,nm device in reverse bias. The dotted lines indicate fits to the linear part of the $IV$-curves, using the expression: $I = R_\text{s}^{-1} V_\text{g} + I_\text{s}$, where $R_\text{s}$ is a serial resistance and $I_\text{s}$ is the saturation current. From thermionic emission theory, we get the saturation current's evolution with temperature, as $I_\text{s} \propto \exp\left(-\Phi_\text{B}/k_\text{B}T\right)$. Fig.~\ref{fig:Sch}.b shows how $\Phi_\text{B}$ can be extracted by fitting such an exponential expression to the $I_\text{s}$ values found in Fig.~\ref{fig:Sch}.a. Finally, Fig.~\ref{fig:Sch}.c shows the determined Schottky barrier heights found for our different device deposition thicknesses. We find generally similar values for the percolation samples, as compared to the bulk sample, with only the 5.5\,nm sample having a slightly lower value of $\Phi_\text{B}$.

%%%%%%%%%%%%%%%%%%%%%%%%%%%%%%%%%%%%%%%%%%%%%%%%%%%%%%%%%%%
\section*{Discussion}
As mentioned previously, the high responsivities observed in the percolation samples versus the bulk film control sample, can likely be explained by two aspects of the percolation geometry. First, the overall thinness of the percolation metal layers in the Schottky junctions helps ensure a higher probability of hole emission (increasing the device's internal quantum efficiency, $\eta_i$). This by itself is due to two major mechanisms: 1 - lower probability of thermalisation due to the nanoscale distance between the location of carrier generation and the Schottky interface, and 2 - the localisation into small metallic grains of $\sim$50\,nm or smaller (see Fig.~1.a) provides an additional avenue of momentum relaxation, which is lost as the feature sizes increase for the thicker deposited films. Finally, due to plasmonic field enhancement and resonant coupling, the metallic part of the junction absorbs a greater deal of incident light.

We see otherwise that responsivity decreases for increasing metal deposition for the percolation structures. This is most likely the result of the metal clusters becoming physically thicker for increasing deposition thickness, lowering the probability of hot carrier emission due to increased transport time/distance, and the nanostructing that contributes to momentum relaxation is decreased for the thicker depositions.

For sensing, we observe a noticeable shift in the overall device responsivity when either an IPA or index matching oil droplet is placed on top of the active diode area, fully embedding the metal structure in the higher index material. Importantly, we also see the diodes recover back to their intrinsic responsivity after the removal of the liquid, demonstrating that the responsivity change is indeed not caused by a chemical change to the metal electrodes. We explain the drop in responsivity as being from a general lowering of optical coupling efficiency to the metal/liquid layer because of a change in the film's effective index, and from lowering the efficiency of any Fabry-P\'erot cavity effect in the samples.

It is worth briefly highlighting how sensing devices such as these could be affected by quantum effects\cite{Jin:2015,Christensen:2017}. It has recently been demonstrated how a high bulk refractive index surrounding plasmonic structures will increase the electron density spill-out, causing additional damping mechanisms of the plasmon resonances to become relevant\cite{Jin:2015}. We believe that the devices investigated here are generally in the classical regime due to the low refractive indices and the low plasmon resonance energies investigated, and as such the plasmonic loss is dominated by phonon/Drude-loss. However, quantum effects such as these could play an important role in future plasmonic index sensing devices, in particular if operating within visible wavelengths.

\section*{Conclusion}
We have demonstrated a simple, cheap, scalable, and CMOS-compatible fabrication technique for Schottky diodes with sub-band gap photodetection in silicon. We utilise self-organsied fractal metasurfaces, known as metal percolation films, to achieve plasmonic enhancement of hot carrier generation across a broad spectral regime. In addition, we have shown how such hot carrier photodetectors can be utilised for bulk refractive index sensing.

Our devices show high responsivities, with a peak value of $\sim$4.5\,mA/W for 1304\,nm wavelength, which is 2 orders of magnitude above that of our bulk film control sample. We find a peak NEP of $\sim$2.70$\cdot10^{-11}$\,W/$\sqrt{\text{Hz}}$ and a corresponding $D^*$ of $\sim$1.48$\cdot10^8$\,Jones, both for 1304\,nm and 0.3\,V reverse bias operation. Such numbers are in the same order of magnitude as some of the commercial photodetectors, e.g. those made of PbS. Our nanostructured devices show internal quantum efficiencies of $\eta_i\sim1.0\%$, which we attribute to momentum relaxation processes made possible by the random structuring and overall thinness of the percolation film geometries. We see generally that the responsivity of the devices decreases with the deposition thickness of the percolation layer, despite the increase in optical absorption. In general we see Schottky barrier heights of $\sim$0.5\,eV, indicating that our devices may be sensitive all the way down to $\sim$2500\,nm wavelengths.

Our devices have no strong dependence on the incident light's polarisation, and show no changes in responsivity for free-space angles of incidence up to $25^\circ$. This makes them a potential candidate for robust and reliable NIR- and SWIR photodetection in many practical applications.

In this work, we have not investigated any ways to optimise the absorption in the aluminium, besides varying the deposition thickness. Future investigations could look into utilising an anti-reflective coating on the illuminated silicon interface, to further increase the transmission of light into the metal. For purely optical photodetection purposes, a thick metallic mirror could be added above the percolation layer, to create a Salisbury screen effect, or even allow plasmonic coupling to the mirror. Such a structure, using gold films, has recently been demonstrated to allow for wideband near-perfect absorption in the visible\cite{Liu:2015} and for contrast enhancement in plasmonic colour printing\cite{Roberts:2018}. Again, we emphasise the relative simplicity of the nanofabrication of our devices, requiring only oxidation of the top silicon surface to define the diode areas, and metal evaporation through a simple mechanical mask while relying purely on CMOS compatible techniques and materials. The fabrication is thus suitable for easily making complex optical photodiode array structures, for applications like NIR-cameras, beam positioning sensors\cite{Grajower:2018b}, and as demonstrated, could also be used for cheap liquid/humidity sensors.

%%%%%%%%%%%%%%%%%%%%%%%%%%%%%%%%%%%%%%%%%%%%%%%%%%%%%%%%%%%
\section*{Methods}
\subsection{Sample fabrication:}
The samples were fabricated from a two-step UV-lithography process. For more details please see Supplementary Fig.~1. 500$\,\upmu$m thick double side polished 4'' wafers of p-doped silicon ($\rho \sim 1$\,$\Upomega$cm) were used. First, a 40\,nm thermal oxide was grown. After this, 150\,nm nitride was grown on the top of the wafer, using plasma-enhanced chemical vapor deposition (PECVD). Next, a pattern for the active areas of the diodes were defined with UV-lithography, and the unmasked nitride was etched with reactive ion etching (RIE). Using local oxidation of silicon (LOCOS), 250\,nm of oxide was grown next to the diode areas. After the LOCOS, the remaining nitride was etched in 180\,$^\circ$C phosphoric acid.

Then, the topside of the wafer is covered with photoresist, and the wafer was baked at 120\,$^\circ$C for 2 minutes. Then the wafer was submerged in buffer hydrofluoric acid (HF) to remove the oxide from the back of the wafer. Then the topside resist is removed by solvent and piranha cleaning.

Next, the wafer is turned over and the any native oxide on the back is removed with HF. Immediately afer, a 100\,nm aluminium film is deposited on the back of the wafer. Using UV-lithography, a pattern for the aluminium Ohmic contact is defined, and the uncovered aluminium is removed with chemical etching. Leftover resist is removed, and the wafer is placed in a 460\,$^\circ$C oven with $95\%/5\%$ N$_2$/H$_2$ atmosphere to make the back Al contact Ohmic by alloying.

Before depositing the thin metal films for the Schottky electrode, the wafer was diced into chips. Each sample is then cleaned, and quickly rinsed in 1:10 HF:H$_2$O, to strip any native oxide at the active diode areas. After this, thin metallic films are deposited with electron-beam evaporation through a mechanical mask. The films are deposited at a rate of 0.6\,\AA/s, in a vacuum chamber pressure of $\sim$10$^{-6}$\,mbar.

\subsection{Responsivity characterisation:}
The samples were characterised using a 1550\,nm Thorlabs fiber coupled laser (model S1FC1550), and a 1304\,nm laser diode powered by a Thorlabs ITC510 Benchtop Laser Diode and Temperature Controller.

A Keysight B2901A Precision Source/Measure Unit was used for recording $IV$-curves. The samples were electrically contacted to the Ohmic back contact with a piece of carbon tape, and a micro-mechanical probe was used to make contact to individual diodes on the top of the sample, by mechanically touching the surrounding metal films.

Light from the corresponding light source was shone through the window of the Ohmic back contact, using an 8\,mm pigtail-style fiber lens from Oz Optics. Using a 3D stage, the light was focused on the individual diode areas, with a spotsize of roughly $\sim 2$\,$\upmu$m diameter. $IV$-characteristics of the diodes were then recorded from $-1.5$\,V to $1.5$\,V, for different wavelengths and powers of illumination. The output power from the fiber lens was measured, and we measured the reflectivity of the silicon substrate to roughly $R=0.4$. The reported incident powers have been corrected as 60\% of the power measured from the fiber lens, to account for the reflection from the bottom of the silicon.

From these measurements, the responsivities were found by fitting the expression:
\begin{align*}
	I_\text{photo}(V_\text{g}) = R(V_\text{g}) P_\text{in} + I_0(V_\text{g}),
\end{align*}
where $I_\text{photo}(V_\text{g})$ is the measured photocurrent for a certain reverse bias of the diode, $V_\text{g}$. $R(V_\text{g})$ is the responsivity for the bias, $P_\text{in}$ in the input optical power, and $I_0(V_\text{g})$ is the measured dark current ($P_\text{in} = 0$\,W) for the same bias. A reverse bias of $V_\text{g} = 0.3$\,V was used for all responsivity calculations.

\subsection{Index sensing:} To measure the effect of different refractive indices on top of the percolation films, the samples were mounted as in the general optical characterisation experiments.  The diodes were then characterised for 5 different optical powers using a 1304\,nm laser diode powered by a Thorlabs ITC510 Benchtop Laser Diode and Temperature Controller. After characterising the dry diode, a liquid droplet of either 95\% IPA or Nikon immersion oil type A was then added, and the diode was measured again for the same 5 optical powers. After each liquid test, the diode was dismounted and washed with acetone and left to dry, and then remounted and remeasured.

\subsection{Schottky barrier measurements:}
The magnitude of the Schottky barriers in the diodes were measured by heating the samples to different temperatures with an Ohmic heater, while recording $IV$-curves from $-1.5$\,V to $1.5$\,V for the different temperatures. From these measurements, the saturation currents from reverse bias, $I_\text{s}(T)$, was found at the different temperatures by fitting the linear parts of the $IV$-curves ($0.3\,\text{V} \leq V_\text{g} \leq 1.5\,\text{V}$) with:
\begin{align*}
	I = R_\text{s}^{-1} V_\text{g} +I_\text{s},
\end{align*}
with $R_\text{s}$ as a serial resistance. The different saturation currents were fitted to the expression\cite{Scales:2010}:
\begin{align*}
	I_\text{s}(T) \propto \exp\left(- \Phi_\text{B} /k_\text{B}T \right),
\end{align*}
where $\Phi_\text{B}$ is the Schottky barrier energy, $k_\text{B}$ is Boltzmann's constant, and $T$ is the temperature.

\subsection{Data availability:} All recorded data are available upon request. The MATLAB code used for the geometric analysis is available at the MATLAB File Exchange: https://www.mathworks.com/\\matlabcentral/fileexchange/71770-fractal-analysis-package

%%%%%%%%%%%%%%%%%%%%%%%%%%%%%%%%%%%%%%%%%%%%%%%%%%%%%%%%%%%
\begin{addendum}
\item[Acknowledgments:] The authors would like to thank Maurice Saidian and Evgenia Blayvas, as well as the rest of the Harvey M. Kruger Family Center of Nanoscience and Technology for assistance in the sample fabrication and SEM characterisation.

This work was financially supported by the Israel-USA Binational science foundation (BSF). 

C.F. is supported by the Carlsberg Foundation as an Internationalisation Fellow.

\item[Author contributions:] The sample geometry was designed by all authors in collaboration. The samples were fabricated by N.M. $IV$-measurements, Schottky barrier measurements, and sensing experiments were done by C.F. with M.G assisting. Absorption measurements were done by J.B.D. and C.F. Image and fractal analysis was done by C.F. The manuscript was written and drafted by C.F., with M.G. and U.L. contributing, and approved by all authors. J.S. and U.L. supervised the entire project.

\item[Conflict of interests:] The authors declare no competing financial interests, or conflicts of interest.

\item[Corresponding authors:] Requests for research materials (data/supplementary information) or general correspondence should be addressed to Christian Frydendahl (\href{christia.frydendahl1@mail.huji.ac.il}{christia.frydendahl1@mail.huji.ac.il}) or Uriel Levy (\href{ulevy@mail.huji.ac.il}{ulevy@mail.huji.ac.il}).

\end{addendum}

\section*{References}
% here the reference list is generated automatically, from the 'references.bib' file

\bibliographystyle{naturemag}
\bibliography{references}

%\bibliographystyle{naturemag}
%\bibliography{references}

%%%%%%%%%%%%%%%%%%%%%%%%%%%ףפ%%%%%%%%%%%%%%%%%%%%%%%%%%%%%%%%

\end{document}